\documentclass[journal=jacsat,manuscript=article]{achemso}

\usepackage{chemformula} 
\usepackage[T1]{fontenc} 
\usepackage{booktabs}
\usepackage{siunitx}
\DeclareSIUnit\angstrom{\text{Å}}
\usepackage[utf8]{inputenc}
\usepackage{epstopdf}
\usepackage{multirow}
\usepackage[hidelinks]{hyperref}
\usepackage{subfig}
\usepackage{xcolor}
\usepackage[version=4]{mhchem}

\author{Irabati Chakraborty}
\affiliation[IIT ISM Dhanbad]
{Department of Physics, Indian Institute of Technology (Indian School of Mines), Atomic and Molecular Physics Laboratory, Dhanbad, Jharkhand 826004, India}
\author{Bobby Antony}
\email{bobby@iitism.ac.in}
\affiliation[IIT ISM Dhanbad]
{Department of Physics, Indian Institute of Technology (Indian School of Mines), Atomic and Molecular Physics Laboratory, Dhanbad, Jharkhand 826004, India}
\title[]
{Investigation of low-energy electron scattering from ethylene glycol}

\abbreviations{}
\keywords{Electron scattering, low energy electron, ethylene glycol, prebiotic molecule, interstellar molecule, R-matrix method, elastic cross section, excitation cross section, differential cross section}

\begin{document}
	
	
	
	\begin{tocentry}
		
		\vfill
		\raisebox{\dimexpr\height-\ht\strutbox+\dp\strutbox\relax}%
		{\includegraphics[width=3.25in, height=1.75in]{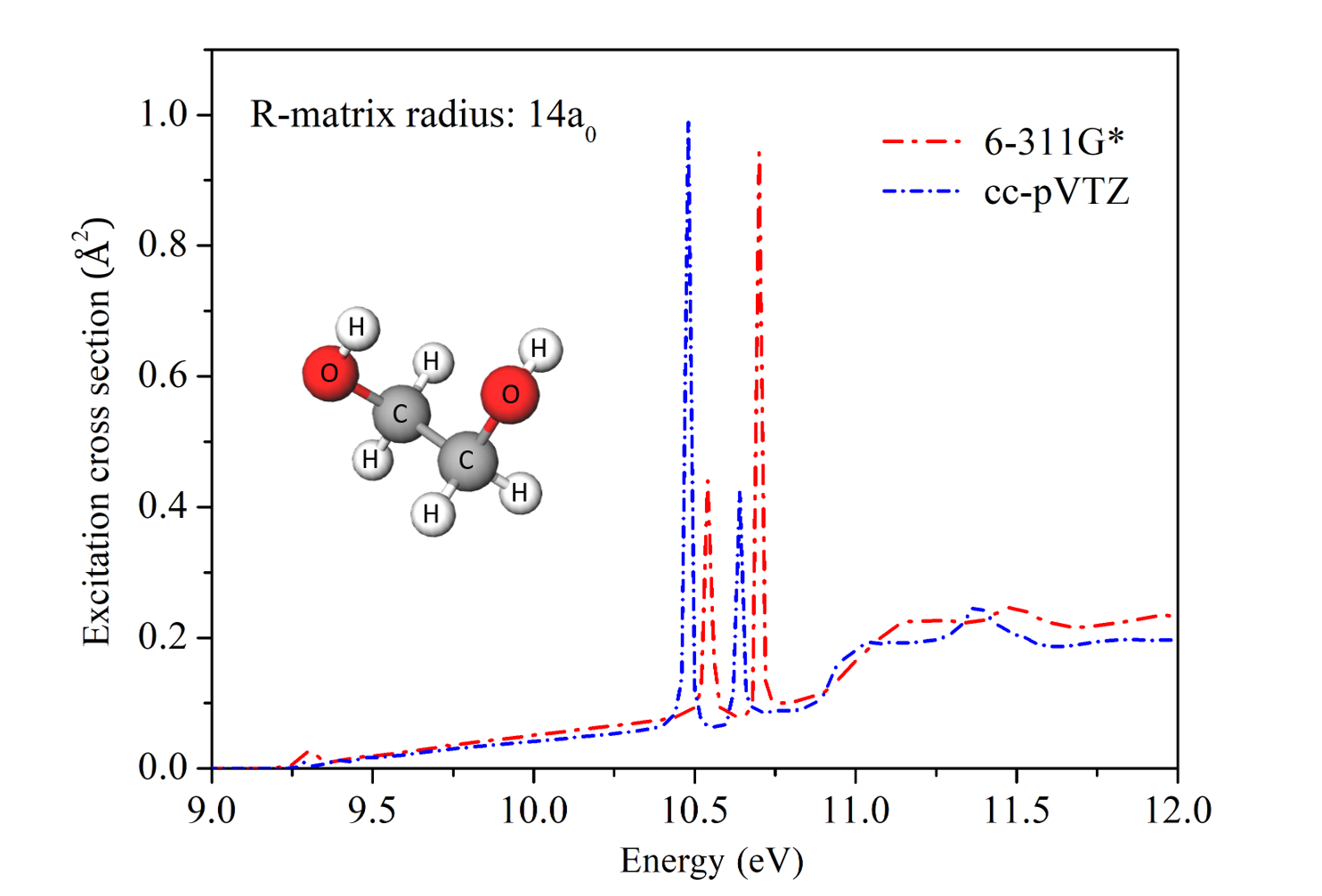}}
		\\\vfill
		
	\end{tocentry}
	
	\begin{abstract}
		Ethylene glycol is a prebiotically relevant complex organic molecule detected in interstellar and cometary environments, yet quantitative low-energy electron-ethylene glycol scattering data remain limited for astrochemical modeling. This work presents an \(R\)-matrix study of low-energy electron collisions with ethylene glycol over the 0 to 12\,eV energy range, using static exchange (SE), static exchange plus polarization (SEP), and configuration interaction (CI) models with 6-311G* and cc-pVTZ basis sets. We compute elastic, excitation, and differential cross sections within a close coupling framework. The dataset offers benchmark inputs for astrochemical models, supporting interpretation of ethylene glycol abundances in space and refining constraints on electron-induced prebiotic pathways.
		
	\end{abstract}
	
	\section{Introduction}
	\label{intro}
	
	Ethylene glycol (EG), \ce{HOCH2CH2OH}, is one of the most significant complex organic molecules (COMs) identified in the interstellar medium (ISM) because of its structural relationship to sugars and its potential role in prebiotic chemistry. Its first confirmed detection was reported in the Galactic Center hot core Sgr B2(N) using millimeter-wave rotational spectroscopy, where Hollis and colleagues identified EG as a reduced alcohol of glycolaldehyde, which is already recognized as a simple sugar-related species \cite{hollis2002interstellar}. Subsequent work extended EG’s interstellar presence to other regions. In Orion-KL, interferometric observations resolved the aGg$^\prime$ conformer \cite{brouillet2015antifreeze}, and later studies identified the gGg$^\prime$ conformer \cite{favre2017complexity}, revealing conformer-dependent spatial distributions within the hot core. In the solar-type protostar IRAS 16293–2422, ALMA surveys revealed rich oxygen-bearing COM chemistry, including glycolaldehyde and ethylene glycol \cite{jorgensen2016alma}, with further studies refining abundances and conformer ratios \cite{nazari2024deep}. More recently, tentative identifications in W51/e2 and G34.3+0.2 \cite{lykke2015tentative} and secure detection in the hot molecular core G358.93–0.03 MM1 \cite{manna2024detection} support the fact that EG is not confined to the Galactic Center but occurs in both high- and low-mass protostellar systems. 
	
	The detection of ethylene glycol in our solar system supports its significance. In comet C/1995 O1 (Hale–Bopp), EG was measured at approximately 0.25\% relative to water, making it one of the most abundant oxygen-bearing organic molecules in that comet \cite{crovisier2004ethylene}. Further detections in comets Lemmon and Lovejoy confirmed EG’s widespread presence in cometary atmospheres at abundances comparable to Hale–Bopp \cite{biver2014complex}, while the Rosetta mission at comet 67P/Churyumov–Gerasimenko also identified EG among other heavy oxygenated species \cite{le2015inventory}. These cometary detections suggest that EG formed in protostellar and protoplanetary environments can survive incorporation into planetesimals and reach young planetary surfaces intact, potentially contributing to the prebiotic organic matter present on the early Earth.
	
	From a chemical and astrobiological perspective, EG is especially important because of its close structural relationship to glycolaldehyde (GA), the simplest sugar, which has been detected toward Sgr B2(N), IRAS 16293–2422, and IRAS2A \cite{jorgensen2012detection, coutens2015detection}. Laboratory experiments indicate that GA and EG often form together in CO/\ce{H2CO}/\ce{CH3OH} ices through radical recombination: HCO + \ce{CH2OH} produces GA, while \ce{CH2OH} + \ce{CH2OH} produces EG \cite{fedoseev2015experimental, butscher2015formation, butscher2016radical}. Such chemistry has been confirmed in controlled hydrogenation experiments at cryogenic temperatures \cite{chuang2016h, simons2020formation, keresztes2024h}. This connection underscores the prebiotic nature of EG as both a tracer and a participant in pathways leading toward sugars \cite{rivilla2017formation}.
	
	\begin{figure}[h]
		\centering
		\includegraphics[scale=1]{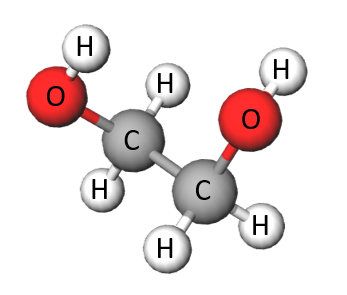}
		\caption{Structure of ethylene glycol}
		\label{fgr:figt1}
	\end{figure}
	
	Low-energy electrons (LEEs), with energies $\leq 20$ eV, have an important influence on the development of EG chemistry in both interstellar and cometary ices. Secondary electrons are produced in large numbers during cosmic-ray ionizations in dense clouds and ices \cite{wu2023role, padovani2018cosmic}. Modeling shows that LEE fluxes in ices can be comparable to or exceed photolysis rates, making them a potentially dominant driver of COM formation under UV-shielded conditions \cite{padovani2018cosmic}. LEEs interact with molecules via resonant scattering processes, forming transient negative ions that can undergo dissociative electron attachment (DEA), dipolar dissociation, or resonant excitation \cite{boyer2016role}. Experiments with methanol-rich ices have demonstrated that electron irradiation yields many of the same products as UV photolysis, including EG itself \cite{boamah2014low, sullivan2016low}. Infrared monitoring of electron-irradiated methanol ices has confirmed formation of GA and EG even with sub-20 eV electrons \cite{sullivan2016low}. These findings highlight the importance of incorporating electron-induced reactions into astrochemical models, as the reactions triggered by low-energy electrons can give rise to distinctive molecular species, which may act as tracers of electron-driven chemistry in the interstellar medium.
	
	Despite the importance of this molecule, as mentioned, low-energy electron scattering data for ethylene glycol remains sparse. Araújo \emph{et al.} computed electron interactions with EG and its isomer dimethyl peroxide, reporting elastic and inelastic cross sections as well as resonance features \cite{araujo2020electron}. Randi \emph{et al.} extended this work with detailed Schwinger multichannel calculations of elastic, differential, and momentum-transfer cross sections, identifying resonances \cite{randi2023methylation}. But these studies have not reported excitation cross sections for electron-EG interaction. In this paper, we present new low-energy electron scattering studies of ethylene glycol to address this gap. Employing the R-matrix method with close-coupling approximations, we provide elastic, excitation, and differential cross sections for electron-EG interaction. These results directly apply to astrochemical kinetics, radiation-driven solid-state chemistry, and the interpretation of conformer- and source-dependent abundance patterns of EG across interstellar and Solar System environments.
	
	\section{Theoretical methodology}
	\label{theory}
	
	The UK R-matrix method \cite{tennyson2010electron}, implemented through the Quantemol-N package \cite{tennyson2007quantemol}, is utilized to determine low-energy electron impact scattering cross sections. The R-matrix method provides a rigorous framework for describing electron-molecule scattering by partitioning configuration space into two regions: an inner region and an outer region, separated by a sphere of radius~$a$, whose center lies at the center of mass of the target molecule. Within the inner region, the target molecule and the incident electron are treated as an $(N+1)$-electron system in which short-range exchange and electron-electron correlation effects are dominant, making the system effectively bound. Beyond this boundary lies the outer region, where the scattering electron interacts only with the long-range multipole potential of the target, and hence short-range forces become negligible. In this asymptotic region, the scattering electron's wavefunction is expanded about a single center, reducing the problem to a set of coupled ordinary differential equations that can be solved efficiently using established numerical techniques. Therefore, the appropriate choice of the radius ($r$) is critically important for guaranteeing that short-range potentials vanish at the boundary. In this work, we have chosen the R-matrix radius to be $14a_0$. The Hamiltonian of the $(N+1)$-electron system is then constructed separately for the inner and outer regions of the R-matrix sphere, in accordance with the relevant interaction terms. These two regions are treated independently, and their solutions are equated at the boundary of the R-matrix sphere to maintain the continuity of the eigenfunctions.
	
	The inner-region wavefunction is obtained using the close-coupling approximation and is given by
	\begin{equation}	
		\psi^{N+1}_k=A\sum_{ij}a_{ijk}\Phi_i(x_1, x_2,....,x_N)u_{ij}({x}_{N+1})+\sum_{i}b_{ik}\chi_i(x_1, x_2,.....,{x}_{N+1}).
		\label{eqn-Inner region-1}
	\end{equation}
	Here $A$ denotes the anti-symmetrization operator, $u_{ij}$ represents the continuum orbitals of the target, and $\Phi_i$ corresponds to the target wavefunctions. The coefficient $a_{ijk}$ denotes the contribution of the product of the $i^{\mathrm{th}}$ target state and the $j^{\mathrm{th}}$ continuum orbital to the $k^{\mathrm{th}}$ inner-region wavefunction. The function $\chi_i$ represents the $L^2$ configuration and is square-integrable, since all electrons occupy orbitals associated with the target wavefunction and are confined within a finite spatial region. The coefficient $b_{ik}$ corresponds to the contribution of the $i^{\mathrm{th}}$ $L^2$ configuration to the $k^{\mathrm{th}}$ inner-region wavefunction.
	
	The eigenfunctions of the inner-region Hamiltonian are employed to construct the R-matrix at the boundary of the inner sphere. This R-matrix is subsequently propagated to the asymptotic region, where it is matched with the asymptotic functions derived from the Gailitis expansion \cite{gailitis1976new}. From this approach, the $K$ matrices are generated, which provide the input for the POLYDCS program \cite{sanna1998differential} to obtain the differential cross sections. The $S$- and $T$-matrices can be derived from the $K$-matrix by applying standard relations. Once the $T$-matrix has been obtained, the corresponding scattering cross sections are evaluated directly from its elements.
	
	To investigate the sensitivity of different theoretical models, from the simpler to more sophisticated ones, calculations were performed using the static exchange (SE), static exchange plus polarization (SEP), and configuration interaction (CI) models. The SE model provides a baseline description in which the target is treated as a frozen Hartree–Fock potential, whereas the SEP model incorporates polarization effects arising from the distortion of the target electronic cloud under the influence of the incident electron. The CI model further refines the description by including excitation processes that are crucial for reproducing resonance features and accurate excitation dynamics.
	
	\subsection{Present Target State}
	
	Ethylene glycol can be in two different point group symmetries: $C_{2h}$ and $C_{1}$. For the present calculation, we have chosen the point group to be $C_{2h}$. The geometrical parameters for ground state ethylene glycol have been taken from the CCCBDB database \cite{CCCBDB}. The basis sets used to construct the target wavefunctions are 6-311G* and cc-pVTZ. The scattering calculations are done using the static exchange (SE), static exchange plus polarization (SEP), and configuration interaction (CI) models. The ground state electronic configuration is given as: 
	\(
	1a_{g}^{2}, \; 1a_{u}^{2}, \; 2a_{g}^{2}, \; 2a_{u}^{2}, \; 3a_{g}^{2}, \; 3a_{u}^{2}, \; 4a_{g}^{2}, \; 4a_{u}^{2}, \; 5a_{u}^{2}, \; 1b_{u}^{2}, \; 5a_{g}^{2}, \; 6a_{g}^{2}, \; 1b_{g}^{2}, \; 6a_{u}^{2}, \; 7a_{g}^{2}, \; 2b_{u}^{2}, \; 2b_{g}^{2}.
	\)
	Additionally, 6 virtual orbitals are included in the CI calculation to represent the electronic states of the molecule beyond the ground state. Out of these 34 electrons, 26 core electrons are frozen and do not take part in the excitation. The remaining 8 electrons in the outermost orbitals take part in the excitation. Hence, the active space is composed of the 4 ground-state outermost orbitals and 6 virtual orbitals, containing 8 electrons, given as: \(
	(7a_{g}, \; 8a_{g}, \; 9a_{g}, \; 6a_{u}, \; 7a_{u}, \; 8a_{u}, \; 2b_{u}, \; 3b_{u}, \; 2b_{g}, \; 3b_{g})^{8}.
	\)
	
	For the CI calculation, a total number of 3540 configuration state functions (CSFs) are generated for the 35 target states.
	
	\section{Results and discussion}
	\label{results}
	This section presents the results obtained for electron-ethylene glycol scattering in the energy range 0 to 12 eV. All the results are presented in graphical formats. 
	
	\subsection{Eigenphase sum}
	
	\begin{figure}[h]
		\centering
		\includegraphics[scale=0.18]{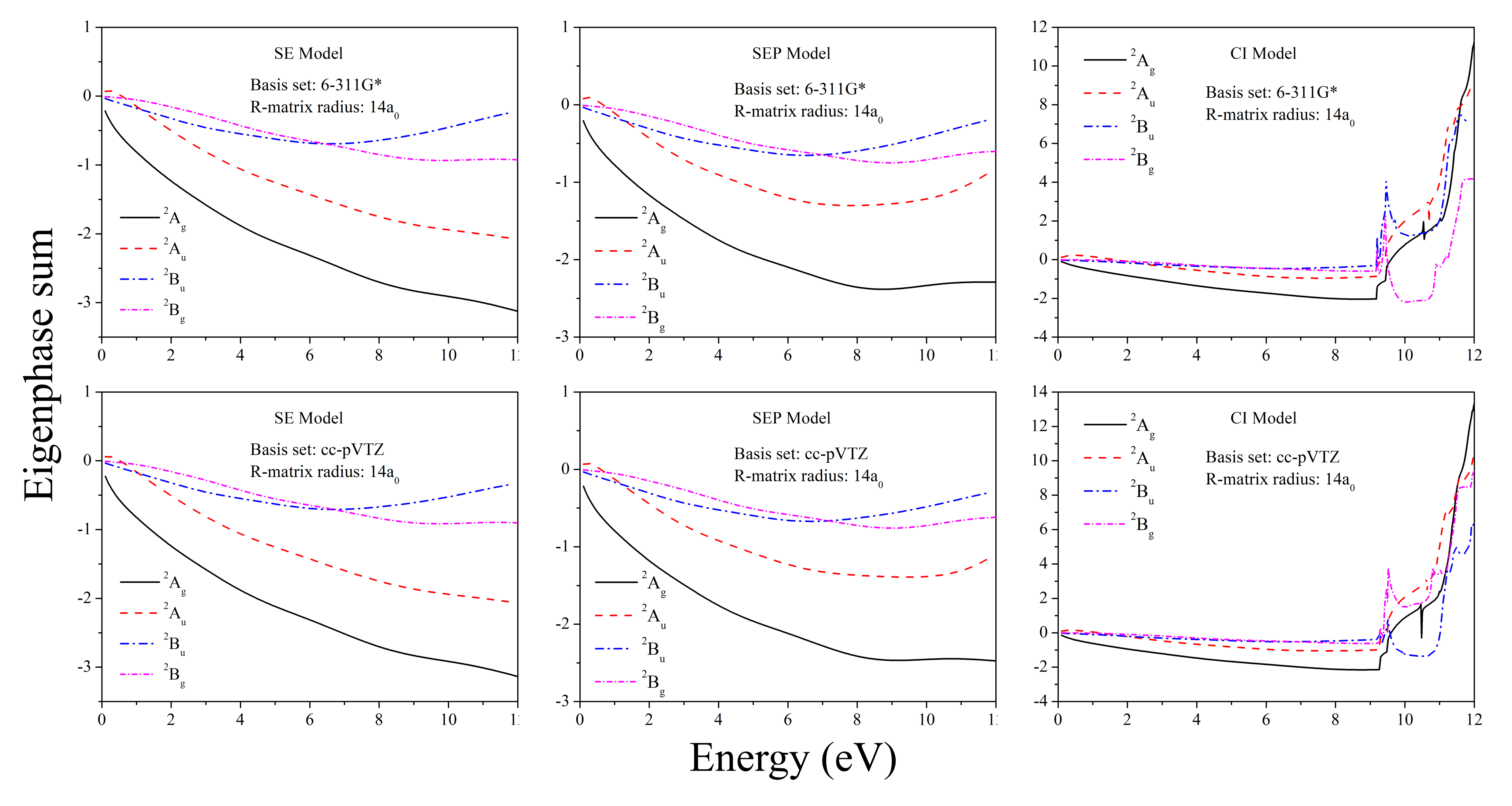}
		\caption{Eigenphase sum of ethylene glycol for different symmetries calculated using SE, SEP, and CI models}
		\label{fig7.1}
	\end{figure}
	
	Figure \ref{fig7.1} represents the eigenphase sums for electron scattering from ethylene glycol, which were calculated for the four symmetry blocks 
	($^{2}A_{g}$, $^{2}A_{u}$, $^{2}B_{u}$, $^{2}B_{g}$) using three levels of target description: static-exchange (SE), static-exchange plus polarization (SEP), and configuration interaction (CI), with two basis sets (6-311G* and cc-pVTZ). In both SE and SEP models, the eigenphase sums vary smoothly with energy, showing monotonic decreases across all symmetries and no sharp $\pi$ jumps, indicating the absence of any resonant states when electron correlation is not explicitly included. Incorporation of polarization in SEP slightly deepens the phase shifts, most notably in the $^{2}A_{g}$ and $^{2}A_{u}$ channels, but remains insufficient to induce resonance features. 
	In contrast, the CI model yields pronounced structures: sharp upward shifts of the eigenphase sums appear near 9–11 eV, with features characteristic of temporary anion formation. Such behavior, appearing only when electronic 
	coupling is included, is consistent with Feshbach-type resonances rather than simple one-electron shape resonances. 
	
	Within the $^{2}A_{g}$ channel, several fitted poles are found between 9.3 and 9.7 eV. The broad features, with widths of about 0.8–1.3 eV, match the gentle curvature observed in the eigenphase trace and can be regarded as 
	genuine short-lived states. A much narrower fitted resonance at 10.55 eV is also reflected in the eigenphase plots with a width of 0.009 eV. When the cc-pVTZ basis is used, the eigenphase structure becomes cleaner and more consistent. 
	In the $^{2}A_{g}$ symmetry, the broad set of overlapping poles collapses into a well-defined narrow resonance at 9.27 eV with a width of about 0.05 eV. Next, in the $^{2}A_{u}$ channel, under the 6-311G* basis set, multiple poles lie between 9.2 and 9.6 eV, with widths of about 0.2–1.1 eV. In contrast, the cc-pVTZ basis produces a clear resonance at 9.27 eV with a much narrower width of 0.008 eV. Another resonance is observed in the 6-311G* basis set at 10.69 eV (width 0.007 eV), which shifts to slightly lower energy in the case of cc-pVTZ, at 10.63 eV (width 0.008 eV). 
	
	In the $^{2}B_{u}$ channel, a resonance is observed at 9.3 eV in the case of the 6-311G* basis set. Although no stable numerical fit was extracted in cc-pVTZ, the sharp $\pi$-like step evident in the eigenphase trace strongly 
	suggests the presence of a Feshbach resonance in this channel. In the $^{2}B_{g}$ channel, a resonance is observed at 9.4 eV for both basis sets. Comparison between basis sets shows that cc-pVTZ produces deeper background phase shifts and slightly lower resonance energies compared with 
	6-311G*, consistent with its more flexible representation of polarization and correlation effects. These resonances are identified as correlation-stabilized Feshbach (core-excited) temporary anion resonances, 
	since they only appear when configuration interaction is included and thus require explicit electron–electron correlation for their formation. Overall, the results demonstrate that while SE and SEP approximations capture only smooth background scattering, inclusion of CI is essential to reproduce the temporary anion resonances of ethylene glycol, with resonance positions clustered in the 9–11 eV range.
	
	\subsection{Elastic cross section}
	
	\begin{figure}[h]
		\centering
		\includegraphics[scale=0.22]{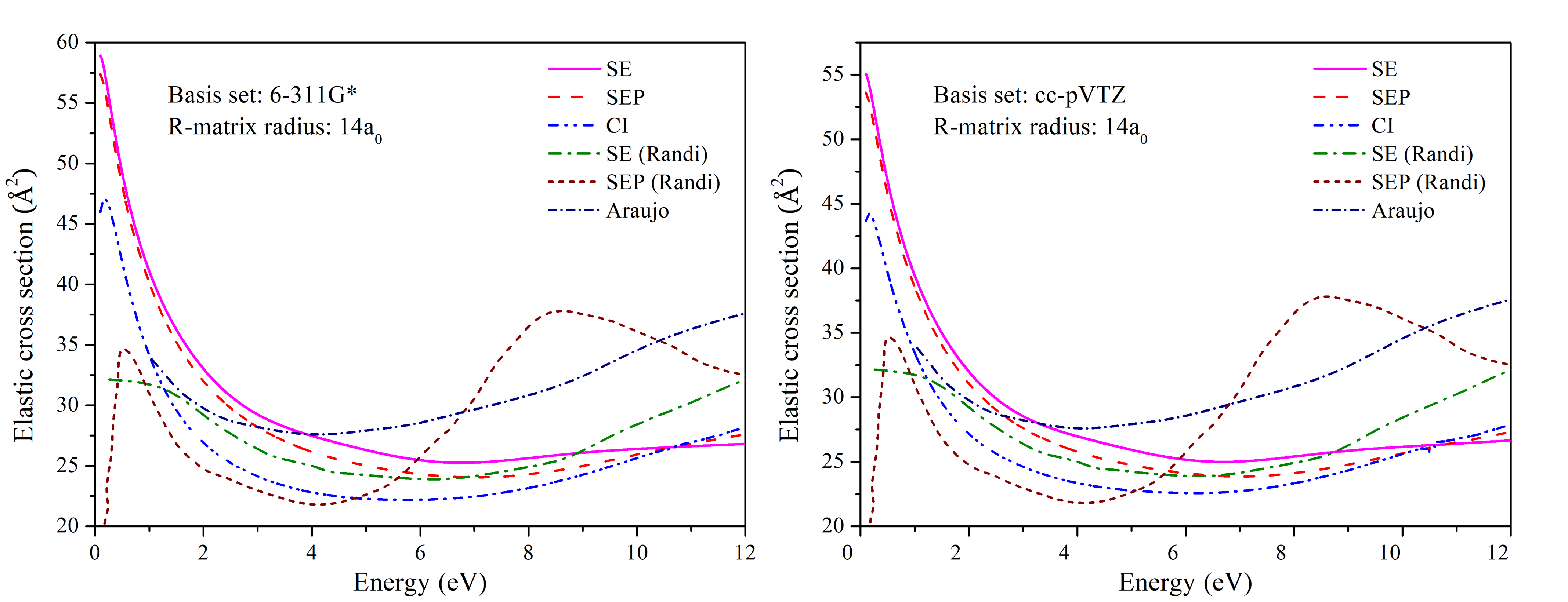}
		\caption{Elastic cross section of ethylene glycol}
		\label{fig7.2}
	\end{figure}
	
	The calculated elastic integral cross sections for electron–ethylene glycol scattering are shown in figure \ref{fig7.2} for the SE, SEP, and CI models using the 6-311G* and cc-pVTZ basis sets, with an $R$-matrix radius of $14 \, a_{0}$. At the threshold the cross section is finite and relatively large. With increasing energy, the cross section decreases rapidly, reaching a broad minimum around 6–7 eV, and then rises again at higher energies. This overall shape is consistently reproduced for both the 6-311G* and cc-pVTZ basis sets, though the absolute magnitudes and fine details differ depending on the scattering model. 
	
	In both basis sets, the SE approximation yields the largest values at threshold, reaching nearly $60 \, \text{\AA}^{2}$ below 1 eV, before decreasing smoothly with energy. This behavior is typical of static-exchange 
	models, which neglect polarization and thus overestimate the scattering probability at low energies. Incorporation of polarization in the SEP model reduces the magnitude of the cross section slightly, bringing the curves into better agreement with earlier theoretical data in the 2–6 eV region. The CI model provides the most refined description. By including coupling to electronically excited states, it lowers the cross sections further in the 5–10 eV range and improves overall agreement with previous calculations. The resonances identified in the eigenphase analysis (9–11 eV) do not manifest as sharp peaks in the integral elastic cross sections. Instead, their effects are subtle and diffused, consistent with narrow Feshbach resonances distributed across several symmetries. As a result, the overall cross-section curves remain smooth. 
	
	Comparison with earlier theoretical work highlights the reliability of the present models. The SMC-SEP result of Randi \emph{et al.} shows a strong structure at low energies and a broad peak around 8–9 eV, which are not observed in our data, suggesting that the SMC method may overestimate polarization effects for this target. The MCOP results of Araujo \emph{et al.}, based on a complex optical potential method, display a steadily increasing cross section above 4 eV, in contrast to the flatter trend observed in the $R$-matrix calculations. Whereas the MCOP cross sections rise to $\sim 35$–$40 \, \text{\AA}^{2}$ by 12 eV, while our cc-pVTZ CI results saturate at $\sim 25$–$27 \, \text{\AA}^{2}$. The cc-pVTZ basis consistently lowers the predicted cross sections relative to 6-311G*, reflecting the improved description of target polarization. Overall, the comparison shows that the present CI model balances the overestimation seen in SE and the stronger energy dependence predicted by the SMC and MCOP methods, delivering a more physically consistent description of elastic scattering.
	
	\subsection{Excitation cross section}
	
	\begin{figure}[h]
		\centering
		\includegraphics[scale=0.18]{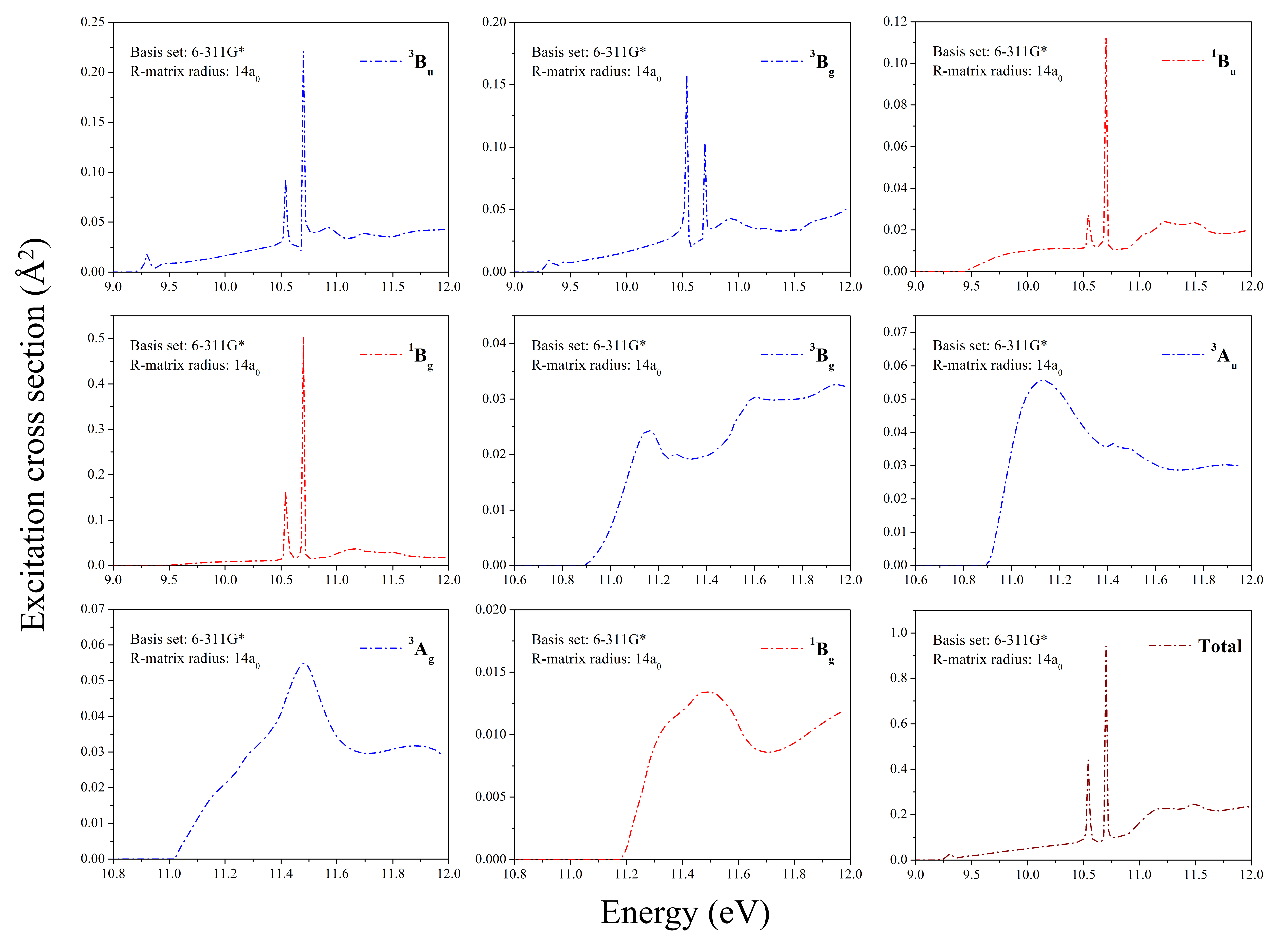}
		\caption{Excitation cross section of ethylene glycol under basis set 6-311G*}
		\label{fig7.3}
	\end{figure}

\begin{figure}[h]
		\centering
		\includegraphics[scale=0.18]{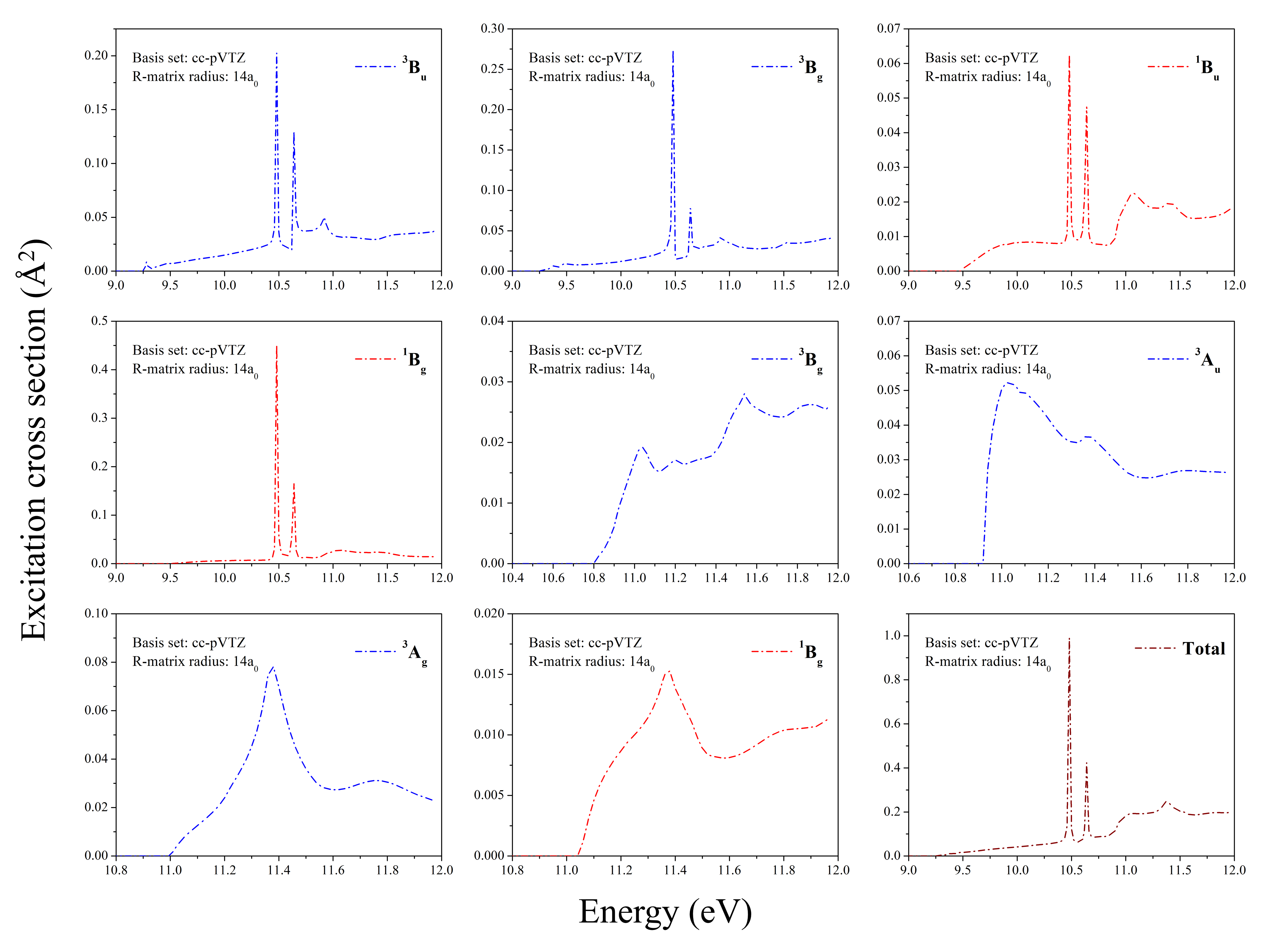}
		\caption{Excitation cross section of ethylene glycol under basis set cc-pVTZ}
		\label{fig7.4}
	\end{figure}

	The integral excitation cross sections ($Q_{\mathrm{EXC}}$) into the lowest singlet and triplet electronic states of ethylene glycol were calculated using the $R$-matrix method with both 6-311G* and cc-pVTZ basis sets, and the results are presented in figures \ref{fig7.3} and \ref{fig7.4}. In both sets of calculations, strong resonance structures are observed in the vicinity of 10.4–10.7 eV, most prominently in the $^{3}B_{u}$, $^{3}B_{g}$, and $^{1}B_{g}$ symmetries. These sharp peaks are characteristic of Feshbach-type resonances, which correspond to the temporary anion states already identified from the eigenphase-sum analysis. Because the $R$-matrix CI model explicitly includes coupling to electronically excited target states, the resonant channels manifest as sharp peaks in the excitation cross sections. 
	In addition to the sharp features near the threshold, smoother structures are seen in the higher-lying triplet excitations. In particular, the $^{3}A_{g}$ and $^{3}A_{u}$ channels dominate above 11.0 eV, with broad maxima centered around 11–11.5 eV. The magnitudes here are modest ($\leq 0.1 \, \text{\AA}^{2}$), but their persistence across both basis sets indicates that these states provide important contributions to the excitation spectrum. 
	
	A comparison between 6-311G* and cc-pVTZ reveals that the overall excitation pattern is robust, but the positions and intensities of individual resonances are slightly basis-dependent. The total excitation cross section, obtained by summing over all channels, is dominated by the two sharp resonances around 10.5 eV and 10.7 eV. In the case of the 6-311G* basis set, the second peak around 10.7 eV has the highest magnitude of $1 \, \text{\AA}^{2}$ and the first peak around 10.5 eV has a magnitude of $0.4 \, \text{\AA}^{2}$. In contrast, under the cc-pVTZ basis set, the first peak 
	around 10.5 eV has the highest magnitude of $1 \, \text{\AA}^{2}$ and the second peak around 10.7 eV has a magnitude of $0.4 \, \text{\AA}^{2}$. Apart from these two prominent peaks, two broad peaks are observed at energies between 11 and 11.5 eV with magnitudes $\leq 0.2 \, \text{\AA}^{2}$. The consistency of the overall profile across both basis sets gives confidence in the robustness of the resonance assignment, even if the precise intensity and width of individual peaks depend somewhat on the basis employed. 
	
	Physically, these results indicate that electron-impact excitation in ethylene glycol is strongly resonance-driven in the 9–11 eV energy region, with the temporary occupation of virtual orbitals coupling most efficiently into the $^{3}B_{u}$, $^{3}B_{g}$, and $^{1}B_{g}$ symmetries. This agrees well with the eigenphase-sum analysis, which identified strong Feshbach resonances across these symmetries in the 9–11 eV energy range.
	
	\subsection{Differential cross section}
	
	\begin{figure}[h]
		\centering
		\includegraphics[scale=0.23]{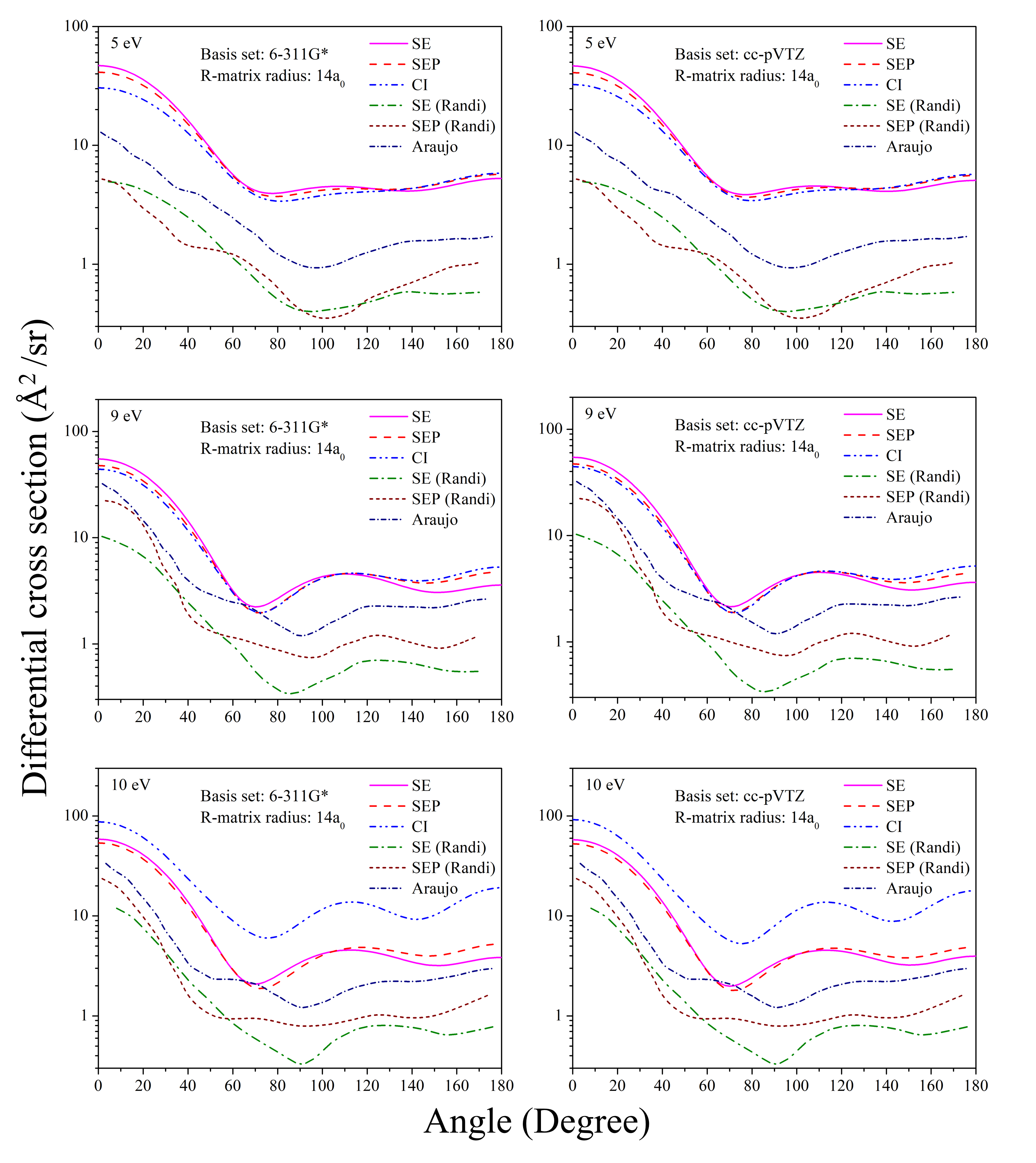}
		\caption{Differential cross section of ethylene glycol at energies 5, 9, and 10 eV}
		\label{fig7.5}
	\end{figure}
	
	The calculated differential cross sections (DCS) for elastic scattering of electrons from ethylene glycol at incident energies of 5, 9, and 10 eV are presented in figure \ref{fig7.5} for both 6-311G* and cc-pVTZ basis sets. The results are compared with theoretical predictions by Randi \emph{et al.} (SMC method) and Araujo \emph{et al.} (MCOP method). In general, our calculated cross sections lie above the available theoretical results across the full angular range and at all three energies considered. In all cases, the angular dependence exhibits the forward-peaked behavior characteristic of electron–molecule scattering, with large values at small scattering angles that decrease steadily toward larger angles. A pronounced minimum is consistently observed near $70^{\circ}$, followed by a secondary rise in the backward direction. This feature reflects interference between different partial-wave contributions and is reproduced across both basis sets and scattering models. 
	
	At 5 eV, beyond the minimum around $70^{\circ}$, the DCS curves are mostly flat for all the models. At 9 eV, the DCS begins to develop a second minimum around $140^{\circ}$–$150^{\circ}$. This second minimum becomes more prominent at 10 eV. At 5 eV and 9 eV, the SE model produces the largest cross sections at forward angles below $40^{\circ}$. Interestingly, at 10 eV, the CI model—which generally reduced the magnitude at lower energies—produces the largest DCS across all scattering angles. This behavior may be linked to the influence of near-threshold Feshbach resonances identified in the eigenphase-sum analysis, whose coupling into the elastic channel enhances the scattering amplitude. Overall, the DCS analysis shows that while the general angular dependence and interference patterns are well 
	reproduced, the absolute magnitudes of our results remain systematically higher than those of Randi \emph{et al.} and Araujo \emph{et al.}

	\section{Conclusion}
	
	In this work, we have presented a comprehensive \(R\)-matrix investigation of electron scattering from ethylene glycol in the energy range 0–12\,eV. By systematically comparing results obtained with static–exchange (SE), static–exchange plus polarization (SEP), and configuration–interaction (CI) target descriptions, and employing two basis sets (6-311G$^{*}$ and cc-pVTZ), we have identified the critical role of electron correlation in shaping the scattering dynamics.
	
	The eigenphase–sum analysis revealed that while SE and SEP models reproduce only smooth, monotonically decreasing background behavior, the CI treatment introduces pronounced resonance structures in the 9–11\,eV region. These resonances, appearing across multiple symmetries and basis sets, are assigned as correlation-stabilized Feshbach-type temporary anion states, highlighting the necessity of including configuration interaction to capture the correct physics of electron attachment in polyatomic alcohols.
	
	Elastic cross sections show the expected rapid falloff from a large threshold value, followed by a broad minimum near 6–7\,eV and a gentle rise at higher energies. The CI results consistently moderate the overestimation of the SE approximation and bring the predictions into improved agreement with earlier studies, while still diverging from SMC and MCOP methods in both shape and magnitude. Importantly, the Feshbach resonances identified in the eigenphase analysis do not appear as sharp structures in the elastic integral cross sections, but their subtle influence is preserved in the scattering amplitudes.
	
	In contrast, the excitation cross sections exhibit sharp and well-defined resonant peaks near 10.4–10.7\,eV, most prominently in the \(^{3}B_{u}\), \(^{3}B_{g}\), and \(^{1}B_{g}\) symmetries, thereby confirming the resonance-driven nature of electron-impact excitation in this system. The robustness of these features across both basis sets supports their assignment to genuine temporary anion states. Additional broad features above 11\,eV point to significant contributions from higher triplet excitations.
	
	Differential cross sections further underline the reliability of the present approach. The angular distributions reproduce the expected forward-scattering dominance and interference minima, while the enhancement observed at 10\,eV in the CI model may be linked to resonance contributions to the elastic channel. Although the absolute magnitudes remain systematically larger than previous theoretical results, the angular trends are internally consistent across models and basis sets.
	
	Overall, this study establishes that electron–ethylene glycol scattering is strongly governed by correlation-induced Feshbach resonances in the 9–11\,eV region, which dominate both excitation and elastic angular distributions. The findings provide new insights into temporary anion formation in polyatomic molecules and underscore the importance of explicitly including electronic correlation for accurate modeling of electron–molecule collisions.

	\begin{acknowledgement}

	\end{acknowledgement}
	
	
	
	\bibliography{ref}
\end{document}